\documentclass{jfm}

\usepackage{graphicx}
\usepackage{natbib}
\usepackage{amsmath}
\usepackage{pifont}
\usepackage{oubraces}
\usepackage{listings}
\usepackage{booktabs}
\usepackage{nicefrac}

\usepackage[normalem]{ulem}

\usepackage{tabularx}
\newcolumntype{Z}{>{\centering\arraybackslash}X}

\usepackage{mathrsfs}
\usepackage{microtype}

\usepackage{mathtools}

\ifCUPmtlplainloaded \else
  \checkfont{eurm10}
  \iffontfound
    \IfFileExists{upmath.sty}
      {\typeout{^^JFound AMS Euler Roman fonts on the system,
                   using the 'upmath' package.^^J}%
       \usepackage{upmath}}
      {\typeout{^^JFound AMS Euler Roman fonts on the system, but you
                   dont seem to have the}%
       \typeout{'upmath' package installed. JFM.cls can take advantage
                 of these fonts,^^Jif you use 'upmath' package.^^J}%
      }
  \else
  \fi
\fi

\ifCUPmtlplainloaded \else
  \checkfont{msam10}
  \iffontfound
    \IfFileExists{amssymb.sty}
      {\typeout{^^JFound AMS Symbol fonts on the system, using the
                'amssymb' package.^^J}%
       \usepackage{amssymb}%
       \let\le=\leqslant  \let\leq=\leqslant
       \let\ge=\geqslant  
      }{}
  \fi
\fi

\ifCUPmtlplainloaded \else
  \IfFileExists{amsbsy.sty}
    {\typeout{^^JFound the 'amsbsy' package on the system, using it.^^J}%
     \usepackage{amsbsy}}
    {}
\fi

\newsavebox{\astrutbox}
\sbox{\astrutbox}{\rule[-5pt]{0pt}{20pt}}

\newcommand\eg{\emph{e.g.}\ }
\newcommand\ie{\emph{i.e.}\ }

\title[A rigid plate on a thin viscous film]
{A pinned or free-floating plate on a thin \\ viscous film. Part 1: A rigid plate}

\author[Trinh, Wilson, and Stone]
{P\ls H\ls I\ls L\ls I\ls P\ls P\ls E\ns H.\ns T\ls R\ls I\ls N\ls H$^1$,\ns
S\ls T\ls E\ls P\ls H\ls E\ls N\ns K.\ns W\ls I\ls L\ls S\ls O\ls N$^2$ \break
\and H\ls O\ls W\ls A\ls R\ls D\ns A.\ns S\ls T\ls O\ls N\ls E$^3$}

\affiliation{
$^1$ Oxford Centre for Industrial and Applied Mathematics (OCIAM), University of Oxford, \\
Mathematical Institute, Andrew Wiles Building, Radcliffe Observatory Quarter, \\
Woodstock Road, Oxford, OX2 6GG, UK \\[\affilskip]
$^2$ Department of Mathematics and Statistics, University of Strathclyde, \\
Livingstone Tower, 26 Richmond Street, Glasgow G1 1XH, UK \\[\affilskip]
$^3$ Department of Mechanical and Aerospace Engineering, Princeton University, \\
Princeton, New Jersey 08544, USA
}

\pubyear{XXXX}
\volume{YYY}
\pagerange{xxx--xxx}
\date{27th November 2013, revised 17th July 2014}

\def\XXint#1#2#3{{\setbox0=\hbox{$#1{#2#3}{\int}$}
     \vcenter{\hbox{$#2#3$}}\kern-.5\wd0}}

\def\de{\,\,\mathrm{d}}
\def\e{\textrm{e}}
\def\tildex{\tilde{x}}
\def\hinf{h_\infty}
\def\xp{x_\text{p}}
\def\hinfzero{h_{\infty 0}}
\def\hinfone{h_{\infty 1}}
\def\cF{\mathcal{F}}
\def\cG{\mathcal{G}}
\def\cH{\mathcal{H}}
\def\pzerocrit{p_0^\text{crit}}
\def\deltastar{\delta^*}
\def\pzerostar{p_0^*}
\def\hinfstar{h_\infty^*}
\def\hhat{\hat{h}}
\def\Xhat{\hat{X}}

\begin{document}

\maketitle

\begin{abstract}
A pinned or free-floating rigid plate lying on the free surface of a thin film of viscous fluid,
which itself lies on top of a horizontal substrate that is moving to the right at a constant speed
is considered.
The focus of the present work is to describe how the competing effects of
the speed of the substrate,
surface tension,
viscosity, and, in the case of a pinned plate,
the prescribed pressure in the reservoir of fluid at its upstream end,
determine the possible equilibrium positions of the plate, the free surface, and the flow within the film.
The present problems are of interest both in their own right as paradigms
for a range of fluid-structure interaction problems in which viscosity and surface tension both
play an important role, and as a first step towards the study of elastic effects.
\end{abstract}

\begin{keywords}
rigid plate,
pinned,
free-floating,
free-boundary problem,
fluid-structure interaction problem,
contact lens
\end{keywords}

\section{Introduction}
\label{sec:intro}

\begin{figure}
\includegraphics[width=1.0\textwidth]{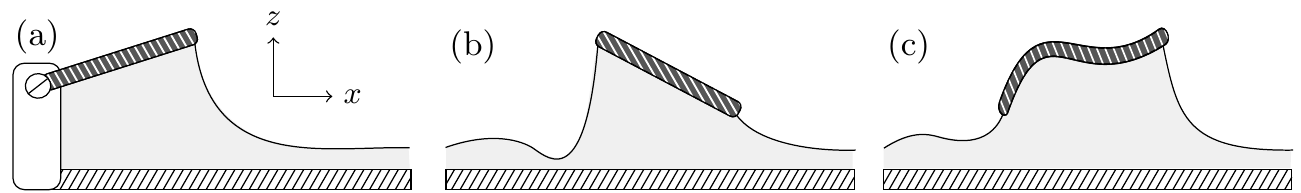}
\caption{
Three paradigm problems involving a rigid or elastic plate lying on the free surface of a film of viscous fluid:
(a) the rigid pinned problem,
(b) the rigid free-floating problem, and
(c) the elastic free-floating problem.
Problems (a) and (b) are studied in the present work.
}
\label{fig:sketch}
\end{figure}

Consider a rigid or elastic plate lying on the free surface of a thin film of viscous fluid, which itself lies on top of a horizontal substrate that is moving to the right at a constant speed.
The plate can either be pinned and in contact with a reservoir of fluid at its upstream end or be free floating, as shown in Figure \ref{fig:sketch}.
Since both the location of the free surface and the position and orientation of the plate are unknown \emph{a priori}, the present problem is both a free-boundary and a fluid-structure interaction problem.
The focus of the present work is to describe how the competing effects of
the speed of the substrate,
surface tension,
viscosity, and, in the case of a pinned plate,
the prescribed pressure in the reservoir,
determine the possible equilibrium positions of the plate, the free surface, and the flow within the film.
%
%
The present work is concerned with the case of a rigid plate, but in Part 2 of our study we will extend our approach to study the influence of the elasticity of the plate.

As an example of the particular challenges that arise in this type of problem even for a rigid plate, consider the work of \cite{moriarty_1996}, who proposed the free-floating problem shown in Figure \ref{fig:sketch}(b) as an idealised model for the motion of a hard contact lens on the tear film of the eye after a blink.
When the substrate moves at a constant speed, the steady-state profile of the free surface exhibits decaying capillary waves upstream and a monotonically decaying profile downstream.
The challenge, as explained by \cite{moriarty_1996} and further discussed in a companion paper by \cite{mcleod_1996}, is coupling the flows upstream and downstream of the plate via the flow underneath it.
These authors demonstrated the existence of either zero, one or two steady-state solutions, depending on the value of an appropriately defined inverse capillary number and the height of the plate relative to the far-field film height.
Subsequently, \cite{quintans_2009} showed that if gravity effects are included then there can be as many as three steady-state solutions.

A limitation of the work by \cite{moriarty_1996} is that they largely ignored the total vertical force and total moment of the forces on the plate and considered the special case of a horizontal plate, \ie the possibility of the plate tilting was mostly discounted.
Concerning the full problem in which the plate is allowed to tilt and both the total vertical force and total moment of the forces on the plate are required to be zero, they state that
\begin{quotation}
\noindent \emph{``In general, it is difficult to find values of $h_0$ [the scaled height of the plate] and $\theta$ [the tilt angle] such that the lens is in equilibrium.''}
\end{quotation}
Furthermore, they conjectured that, for general values of the capillary number,
there is a height at which a horizontal plate is in equilibrium.
In the present work we shall demonstrate that this conjecture is not, in fact, true by showing that the horizontal configuration is only an equilibrium solution in the two extreme cases of zero and infinite capillary numbers; for general values of the capillary number, there is an equilibrium solution with non-zero tilt angle.

It should be noted that \cite{moriarty_1996} were not the first to consider the free-floating problem as a model for contact lens motion, and others (\eg \citealt{conway_1983} and \citealt{kamiyama_2000}) have considered similar configurations in the context of lens modelling.
Moreover, there is a large body of literature on the subject of modelling contact lenses including additional physical effects, such as tear film rupture (\eg \citealt{wong_1996}), porosity of the lens (\eg \citealt{nong_2010}), and the suction pressure underneath the lens (\eg \citealt{maki_2014}).

Our strategy for studying the free-floating problem shown in Figure \ref{fig:sketch}(b) is to first consider the pinned problem shown in Figure \ref{fig:sketch}(a).
The plate is pinned at a fixed location and in contact with a reservoir of fluid at its upstream end, and coupled to a single free surface at its downstream end.
In addition to reducing the complexity of the calculations, the pinned problem is of interest in its own right (see below).
Furthermore, as we shall see, once the analytical and numerical subtleties of the pinned problem are understood, then the extension to the free-floating problem is relatively straightforward (at least in principle, if not always in practice).

\begin{table}
\begin{center}
\begin{tabular}{cccc}
& Free Surfaces & Surface Tension & Elasticity \\
\cite{conway_1983}        & $0$        & no  & yes \\
\cite{moriarty_1996}      & $2$        & yes & no  \\
\cite{kamiyama_2000}      & $0$        & no  & yes \\
\cite{hosoi_2004}         & $0$        & no  & yes \\
\cite{giacomin_2012}      & $1$        & no  & yes \\
\cite{hewitt_2012b}       & $0$        & no  & no  \\
\cite{dixit_2013}         & $0$        & no  & yes \\
This work (Parts 1 and 2) & $1$ or $2$ & yes & no and yes \\
\end{tabular}
\end{center}
\caption{
A selection of previous work on both the pinned and free-floating problems.
}
\label{tab:res}
\end{table}

Table \ref{tab:res} shows a selection of previous work on both the pinned and free-floating problems for both rigid and elastic plates, and illustrates that the much of the previous work has ignored the effect of surface tension, which is a key ingredient of the present work.
The present work is concerned with the case of a rigid plate,
but in Part 2 of our study we will analyse the case of an elastic plate.

While the present problems are of interest in their own right as paradigms for a range of fluid-structure interaction problems in which viscosity and surface tension both play an important role, the present study was directly inspired by the recent growth of both experimental and theoretical interest in a variety of elastocapillary problems. 
These include
the wetting of fibrous material
(\eg \citealt{bico_2004}, \citealt{duprat_2012}, \citealt{taroni_2012}, and \citealt{singh_2014}),
the buckling of floating elastic sheets
(\eg \citealt{hosoi_2004}, \citealt{audoly_2011}, and \citealt{wagner_2011}),
blade coating using a flexible blade (\eg \citealt{pranckh_1990}, \citealt{iliopoulos_2005}, and \citealt{giacomin_2012}),
and
the elastic drag-out problem
(\eg \citealt{dixit_2013}).
Also relevant here is the recent work by \cite{hewitt_2012a} and \cite{hewitt_2012b} on the so-called ``washboard'' instability which can be created by moving a pivoted rigid plate over a layer of fluid or granular material.
While in many situations the effect of elasticity plays a key role, one of the main conclusions of the present
work is that even the deceptively simple problem of a rigid plate moving relative to a rigid substrate has considerable
difficulties and subtleties which must be understood before elastic effects can be fully understood.

The outline of our paper is as follows.
In \S\ref{sec:mathform}, we formulate the equations and boundary conditions for a rigid plate lying on the free surface of a thin film of viscous fluid.
Asymptotic and numerical analyses of the pinned problem and of the free-floating problem are described in \S\ref{sec:pinned} and \S\ref{sec:free}, respectively.
Finally in \S\ref{sec:discussion}, we conclude with a short discussion.

\section{Mathematical formulation for a rigid plate}
\label{sec:mathform}

Consider the steady two-dimensional flow of a thin film of Newtonian fluid with constant density $\rho$, viscosity $\mu$, and surface tension $\gamma$, which lies on top of a rigid horizontal substrate that is moving to the right with constant speed $U$ and is located at $z = 0$ in the natural Cartesian coordinate system $(x,z)$.
A rigid plate of length $L$ lies on the free surface of the film, which is denoted by $z = h(x)$, and
the height of the plate above the substrate is denoted by $z = H(x)$ for $0 < x < L$.
Note that while the shape of the plate is prescribed, both its position and orientation are unknown, and hence, like the location of the free surface, the location of the plate has to be determined as part of the solution to the problem.
For the present, the shape of the plate will be left general, but subsequently we shall consider the case of a flat (but not, in general, horizontal) plate.

While the following description of the thin-film flow underneath the plate and the free surface is now relatively standard,
what makes the present problem challenging is the freedom in both the position and orientation of the plate, and
the coupling of the regions upstream and downstream of the plate via the flow underneath it.

The pressure and velocity of the fluid are denoted by $p=p(x)$ and $\textbf{u} = (u(x,z),w(x,z))$, respectively, and we assume that the atmosphere above the plate and the free surface is inviscid and at a uniform atmospheric pressure which we take to be zero without loss of generality.
Since the systems we are interested in are typically small, the effect of gravity will be neglected in the present analysis.
Using primes $(')$ to denote non-dimensional quantities, we non-dimensionalise the variables according to
\begin{equation}
x = L x', \quad
z = \epsilon L z', \quad
h = \epsilon L h', \quad
u = U u', \quad
v = \epsilon U v', \quad
p = \frac{\mu U}{\epsilon^2 L} p',
\label{nondimen}
\end{equation}
where $\epsilon = \cH/L \ll 1$ is the aspect ratio of the film in which the characteristic dimensional film height $\cH$ will be defined explicitly for the two problems we will consider in \S\ref{sec:pinned} and \S\ref{sec:free}, respectively.
Now dropping the prime notation and assuming that all variables are henceforth dimensionless unless stated otherwise, we find that at leading order in the thin-film (\ie small aspect ratio) limit, $\epsilon \to 0$, the mass conservation and Navier--Stokes equations reduce to the familiar two-dimensional lubrication equations, given by
\begin{subequations}
\begin{alignat}{3}
\text{continuity:} \quad &
u_x + w_z = 0, &
\label{lubeqns1} \\
\text{$x$-momentum:} \quad &
p_x = u_{zz}, &
\label{lubeqns2} \\
\text{$z$-momentum:} \quad &
p_z = 0, &
\label{lubeqns3}
\end{alignat}
\end{subequations}
where subscripts denote derivatives.
On the substrate $z = 0$ and the plate $z = H(x)$
the usual no-slip and no-penetration conditions apply, whereas
on the (clean) free surface $z = h(x)$
the usual balances of normal and tangential stress hold.
In the thin-film limit the appropriate boundary conditions are
\begin{subequations}
\begin{alignat}{4}
\text{no slip and no penetration on the substrate:} \quad &
(u,w) = (1,0) \quad & &
\text{on} \quad z = 0,
\label{noslipsubstrate} \\
\text{no slip and no penetration on the plate:} \quad &
(u,w) = (0,0) \quad & &
\text{on} \quad z = H(x),
\label{noslipplate} \\
\text{normal stress on the free surface:} \quad &
p = -\delta^3 h_{xx} \quad & &
\text{on} \quad z = h(x),
\label{normal} \\
\text{tangential stress on the free surface:} \quad &
u_z = 0 \quad & &
\text{on} \quad z = h(x),
\label{tangential}
\end{alignat}
\end{subequations}
where
\begin{equation}
\delta^3 = \frac{\gamma \epsilon^3}{\mu U} = \frac{1}{\textrm{Ca}}
\end{equation}
is a suitably defined inverse capillary number.
(Note that this definition of $\delta$ differs by a factor of 3 from that used by \citealt{moriarty_1996}.)

\subsection{Flow underneath the free surface}

Underneath the free surface, the momentum equations (\ref{lubeqns2}) and (\ref{lubeqns3}) can be integrated, and using the boundary conditions (\ref{noslipsubstrate}) and (\ref{tangential}) yields the horizontal velocity
\begin{equation}
u = \frac{p_x}{2}(z^2-2hz)+1.
\label{u_freesurface}
\end{equation}

The constant volume flux per unit width (non-dimensionalised with $\epsilon U L$) underneath the free surface, $Q$, is therefore given by
\begin{equation}
Q = \int_0^h u \de{z} = -\frac{h^3 p_x}{3}+h,
\label{Q_freesurface}
\end{equation}
and hence the pressure $p$ satisfies
\begin{equation}
p_x = \frac{3(h-Q)}{h^3}.
\label{px_freesurface}
\end{equation}

Substituting for $p_x$
from (\ref{px_freesurface}) into (\ref{u_freesurface})
and re-arranging yields
\begin{equation}
u = \frac{3(h-Q)(z-2h)z+2h^3}{2h^3}.
\label{u_freesurface_Q}
\end{equation}

Since from (\ref{lubeqns3}) $p$ is independent of $z$,
then the boundary condition (\ref{normal}) means that
\begin{equation}
p=-\delta^3 h_{xx}
\label{p_freesurface}
\end{equation}
throughout the fluid, and so substituting for $p$ into (\ref{px_freesurface}) shows that the free surface $h$ satisfies the familiar Landau--Levich equation
\begin{equation}
\delta^3 h_{xxx} = \frac{3(Q-h)}{h^3}
\label{LandauLevich}
\end{equation}
(see, for example, \citealt{tuck_1990}).
Far from the plate the free surface approaches the uniform far-field height $\hinf$,
\ie $h \to \hinf$, and hence from (\ref{LandauLevich}) the flux is given by $Q=\hinf$.

\subsection{Flow underneath the plate}

Underneath the plate (\ie for $0 \leq x \leq 1$), the momentum equations (\ref{lubeqns2}) and (\ref{lubeqns3}) can again be integrated, and using the boundary conditions (\ref{noslipsubstrate}) and (\ref{noslipplate}) yields the horizontal velocity
\begin{equation}
u = \frac{p_x}{2}(z^2-Hz)-\frac{z}{H}+1.
\label{u_plate}
\end{equation}

The flux underneath the plate (which must, of course, be equal to the flux underneath the free surface given by (\ref{Q_freesurface})), $Q$, is therefore given by
\begin{equation}
Q = \int_0^H u \de{z} = -\frac{H^3 p_x}{12}+\frac{H}{2},
\label{Q_plate}
\end{equation}
and hence the pressure $p$ satisfies
\begin{equation}
p_x = \frac{6(H-2Q)}{H^3}.
\label{px_plate}
\end{equation}

Substituting for $p_x$
from (\ref{px_plate}) into (\ref{u_plate})
and re-arranging yields
\begin{equation}
u = \frac{[3(2Q-H)z+H^2](H-z)}{H^3}.
\label{u_plate_Q}
\end{equation}

Integrating (\ref{px_plate}) with respect to $x$
shows that $p$ is given by
\begin{equation}
p = p(0) + 6I_2(x) - 12QI_3(x),
\label{p_plate}
\end{equation}
where, for convenience, we have defined the integral $I_n$ by
\begin{equation}
I_n(x) = \int_0^x \frac{1}{H^n(\tildex)} \de{\tildex}
\label{In}
\end{equation}
(see, for example, \citealt{duffy_2007}).

\subsection{A rigid flat plate}

Thus far, the formulation of the problem has been for a rigid plate of general shape.
For simplicity, we now restrict our attention to the particular choice of a flat (but not, in general, horizontal) plate given by
\begin{equation}
H = H_0+\alpha x,
\label{plateeq}
\end{equation}
where $\alpha$ (which must satisfy $\alpha > -H_0$ for the solution to be physically realisable) is the unknown tilt angle measured anti-clockwise from the horizontal.

In this case, the integrals $I_2$ and $I_3$ defined by (\ref{In}) can be evaluated readily, and, for convenience, are given by (\ref{I2}) and (\ref{I3}) in the Appendix. Hence from (\ref{p_plate}) the pressure underneath the plate is given by
\begin{equation}
p = p(0)+\frac{6\left[H_0(H_0+\alpha x)-Q(2H_0+\alpha x)\right]x}{H_0^2(H_0+\alpha x)^2}.
\label{pressure1}
\end{equation}
Note that (\ref{pressure1}) is simply the classical solution for the pressure underneath a linear slider bearing, and setting $p(1)=p(0)$ would recover the classical result for the flux, namely $Q=H_0(H_0+\alpha)/(2H_0+\alpha)$.
However, in the present problem the value of $Q=\hinf$ is determined by the far-field film height and the situation is considerably more complicated due to the coupling of the pressures at the two ends of the plate via the flow underneath it, the influence of surface tension, and the freedom in both the position and orientation of the plate.

\section{The pinned problem}
\label{sec:pinned}

Consider now the case of flow underneath a rigid flat plate that is pinned at a fixed dimensional height $H_0$ and in contact with a reservoir of fluid held at a prescribed constant dimensional pressure $P_0$ at its upstream end, as shown in Figure \ref{fig:sketch}(a).

In this problem we choose $\cH=H_0$ for the characteristic dimensional film height in (\ref{nondimen});
this sets the non-dimensional height of the upstream (\ie the pinned) end of the plate to be $H(0)=1$, and so from (\ref{plateeq}) the non-dimensional height of the plate is given by $H=1+\alpha x$, where $\alpha > -1$, and the non-dimensional prescribed pressure is denoted by $p_0=p(0)$.
The solutions for the velocity $u$ and pressure $p$ underneath the free surface are given by (\ref{u_freesurface_Q}) and (\ref{p_freesurface}), while the corresponding solutions underneath the plate are given by (\ref{u_plate_Q}) and (\ref{p_plate}).
The height of the free surface downstream of the plate is obtained by solving the Landau--Levich equation (\ref{LandauLevich}) subject to appropriate matching conditions at the downstream (\ie the free) end of the plate and the far-field condition $h \to \hinf$ as $x \to \infty$.

\begin{table}
\begin{center}
\begin{tabular}{cp{4cm}p{6cm}}
\qquad Unknowns \qquad \qquad & Equations & Boundary conditions \\
$\alpha$, $\hinf$ & $H=1+\alpha x$
\newline $\delta^3 h_{xxx}=3(\hinf-h)/h^3$
\newline $p_x=6(H-2\hinf)/H^3$ & [1] $p(0)=p_0$
\newline [2] $h(1)=H(1)=1+\alpha$
\newline [3] $[p]=0$ at $x=1$
\newline [4,5] $h \to \hinf$ as $x \to \infty$
\newline [6] $M_0=0$ \\
\end{tabular}
\end{center}
\caption{
A summary of the equations and boundary conditions of the sixth-order rigid pinned problem.
}
\label{tab:pinned}
\end{table}

\begin{figure}
\includegraphics[width=1.0\textwidth]{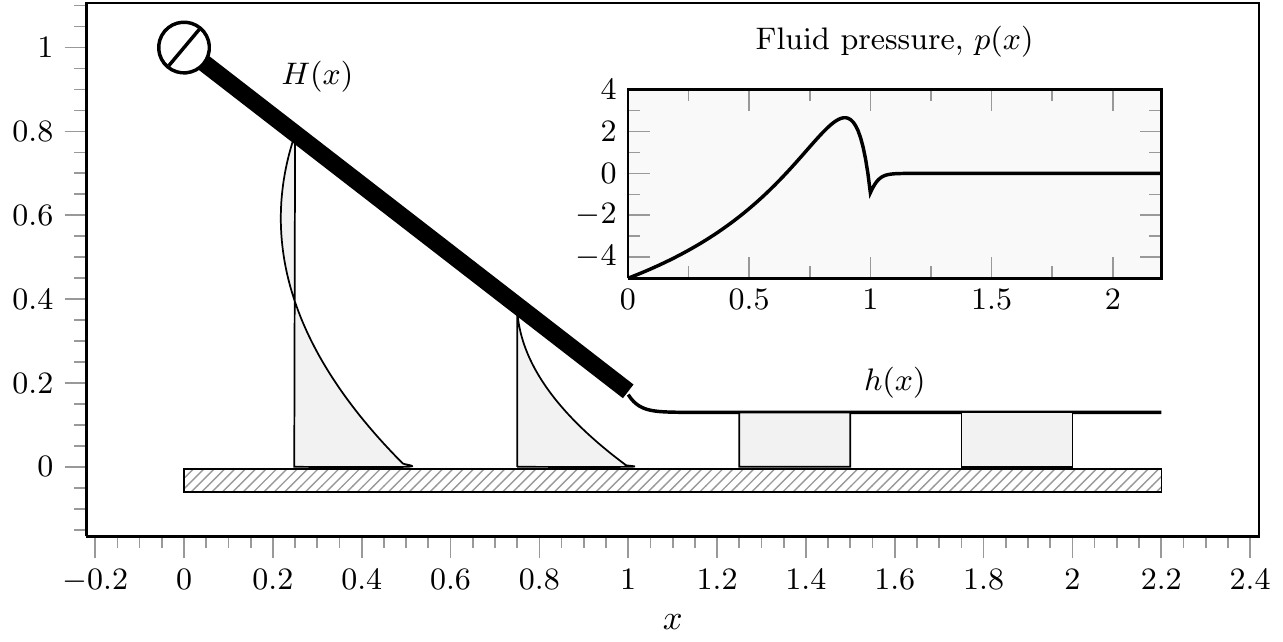}
\caption{
The solution of the pinned problem when $\delta = 0.25$ and $p_0=-5$, for which we find numerically that $\alpha \simeq -0.828$ and $\hinf \simeq 0.130$. Re-scaled velocity profiles, $0.25 \, u(x,z)$, are plotted as functions of $z$ for $x = 0.25$, 0.75, 1.25, 1.75. The inset shows the fluid pressure, $p(x)$.
}
\label{fig:pinned}
\end{figure}

A summary of all the relevant equations and boundary conditions is given in Table \ref{tab:pinned}, and the solution of the pinned problem when $\delta=0.25$ and $p_0=-5$, computed numerically as described subsequently in \S\ref{sec:num1}, is shown in Figure \ref{fig:pinned}.
We note that in Table \ref{tab:pinned} the downstream condition provides two boundary conditions.
The reason for this is that if we write $h = \hinf + \overline{h}$, where $\overline{h} \ll \hinf$, then according to a standard WKB analysis,
\begin{equation}
\overline{h} \sim
  C_1 \exp \left( -\frac{3^{1/3} x}{\delta\hinf} \right)
+ C_2 \exp \left(  \frac{3^{1/3}\e^{ \pi i/3} x}{\delta\hinf} \right)
+ C_3 \exp \left(  \frac{3^{1/3}\e^{-\pi i/3} x}{\delta\hinf} \right)
\label{eq:wkbmodes}
\end{equation}
as $x \to \infty$ (see, for example, \citealt{tuck_1990}).
The two exponentially growing modes, which represent capillary waves, are ruled out on physical grounds, so that $C_2=0$ and $C_3=0$, and leaving only $C_1$ to be determined.

At the downstream end of the plate we assume that the free surface is pinned to the end of plate and
impose continuity of height, namely $h(1)=H(1)=1+\alpha$, and pressure.
The pressure of the fluid underneath the plate (\ref{p_plate}) is continuous with the pressure of the fluid underneath the free surface (\ref{p_freesurface}) at $x=1$ provided that
\begin{equation}
p_0 + 6I_2(1) - 12\hinf I_3(1) = - \delta^3 h_{xx}(1),
\label{p1}
\end{equation}
and hence
\begin{equation}
p_0 + \frac{6\left[1+\alpha-\hinf(2+\alpha)\right]}{(1+\alpha)^2}
+ \delta^3 h_{xx}(1) = 0.
\label{p2}
\end{equation}
The condition (\ref{p2}) explicitly couples the free surface downstream of the plate to the prescribed pressure at its upstream end, $p_0$, via the flow underneath it.

The final condition requires zero total moment of the forces on the plate about its pinned end.
The expression for the moment per unit width (non-dimensionalised with $\mu U L / \epsilon^2$) about $x=0$, $M_0$, is given by
\begin{equation}
M_0 = \int_0^1 x p \de{x}
+ \delta^3 [h_x(1)+H(0)-H(1)] = 0,
\label{M1}
\end{equation}
where the first term represents the moment of the normal force on the plate due to the fluid underneath it, and the second term is the moment of the normal surface-tension force acting at the downstream end of the plate, $x=1$.
Substituting the expression for the pressure underneath the plate (\ref{p_plate}) into (\ref{M1}) yields
\begin{equation}
M_0 = \frac{p_0}{2}
+ \int_0^1 \left[ 6 x I_2(x) - 12 \hinf x I_3(x) \right] \de{x}
+ \delta^3 [h_x(1)-\alpha] = 0,
\label{M2}
\end{equation}
where the integrals of $x I_2$ and $x I_3$ (both of which are functions of $\alpha$) are given by (\ref{xI2}) and (\ref{xI3}) in the Appendix with $H_0=1$.

The two key unknowns in the pinned problem are thus the tilt angle, $\alpha$, and the far-field film height, $\hinf$; once these two quantities are determined, the problem is solved for a given $p_0$ and $\delta$. The counting argument in Table \ref{tab:pinned} shows that the pinned problem is a sixth-order problem, which is completely determined through the six boundary conditions we have just discussed.

The horizontal and vertical components of the total force per unit width (non-dimensionalised with $\mu U / \epsilon^2$) on the plate, $\mathbf{F} = (F_x, F_z)$, are given by
\begin{subequations}
\begin{align}
F_x &= \frac{\delta^3}{\epsilon} - \epsilon \left[ \int_0^1 H_x p \de{x} + \int_0^1 u_z \Bigr\rvert_{z = H} \de{x} + \frac{\delta^3}{2} h_x^2(1) \right], \label{Fx_pinned} \\
F_z &= \int_0^1 p \de{x} + \delta^3 h_x(1),
\end{align}
\end{subequations}
respectively.
Using the expressions for the velocity underneath the plate (\ref{u_plate_Q}), the height of the plate (\ref{plateeq}), the pressure underneath the plate (\ref{pressure1}), and the integrals of $I_2$ and $I_3$ given by (\ref{I2int}) and (\ref{I3int}) in the Appendix, we obtain
\begin{subequations}
\begin{align}
F_x &= \frac{\delta^3}{\epsilon} - \epsilon\left[ \alpha p_0 + \frac{2\{ 3\alpha - 2\log(1+\alpha)\}}{\alpha} - 6\hinf + \frac{\delta^3}{2} h_x^2(1) \right], \\
F_z &= p_0 + \frac{6\{ \alpha - \log(1+\alpha)\}}{\alpha^2} - \frac{6\hinf}{1+\alpha} + \delta^3 h_x(1).
\end{align}
\end{subequations}
In the present case of a pinned plate, the pivot will exert the necessary reactive force such that the horizontal and vertical forces are balanced; however, for the case of a free-floating plate, the force balances require further consideration, which will be discussed in \S\ref{sec:realistic}.

\subsection{Numerical method for the pinned problem}
\label{sec:num1}

For general values of $\delta$, the pinned problem can be solved numerically by reposing the boundary value problem summarised in Table \ref{tab:pinned} as the following nested shooting problem (see, for example, \citealt{tuck_1990}).

For each value of $\delta$, begin by guessing an initial value of the tilt angle,
\begin{equation}
\alpha \simeq \alpha_\text{guess},
\end{equation}
and the far-field film height,
\begin{equation}
\hinf \simeq h_{\infty,\text{guess}}.
\end{equation}

Next, by writing $h = h_{\infty,\text{guess}} + \kappa$, where $\kappa$ is a numerically small perturbation ($\kappa = 10^{-8}$ in our calculations), the far-field behaviour (\ref{eq:wkbmodes}) is used to provide the values of $h'$ and $h''$. With these three initial conditions, the Landau--Levich equation (\ref{LandauLevich}) is solved for the film height, $h$, until reaching the value of $h = 1 + \alpha_\text{guess}$ at the downstream end of the plate.

At this point, the pressure and surface-tension forces at the ends of the plate are known (albeit for incorrect values of $\alpha$ and $\hinf$). The correct value of $\alpha$ is found by solving the zero moment condition (\ref{M2}) using Newton's method. Finally, the correct value of $\hinf$ is found by requiring that the pressure at $x=0$ is equal to the prescribed pressure, \ie $p(0) = p_0$. This nested shooting method can be concisely written as the solution of the pair of coupled equations,
\begin{equation}
\cF(\hinf; \, \delta) \equiv p(0) - p_0 = 0
\qquad \text{and} \qquad
\cG(\alpha; \, \hinf, \delta) \equiv M_0 = 0.
\end{equation}

In general, we have found that the pinned problem is robust to most choices for the initial conditions, \ie most guesses for $\alpha$ and $\hinf$ will lead to convergence. Solutions for a range of values in $\delta$ can be computed by beginning with either a very small or a very large value of $\delta$, and using the asymptotic solutions obtained subsequently in \S\ref{sec:pinned_largedelt} and \S\ref{sec:pinned_smalldelt} as initial guesses. Solutions for other values of $\delta$ are then computed through numerical continuation. The solution of the pinned problem when $\delta=0.25$ and $p_0=-5$ was shown in Figure \ref{fig:pinned}.

\begin{figure}
\includegraphics[width=1.0\textwidth]{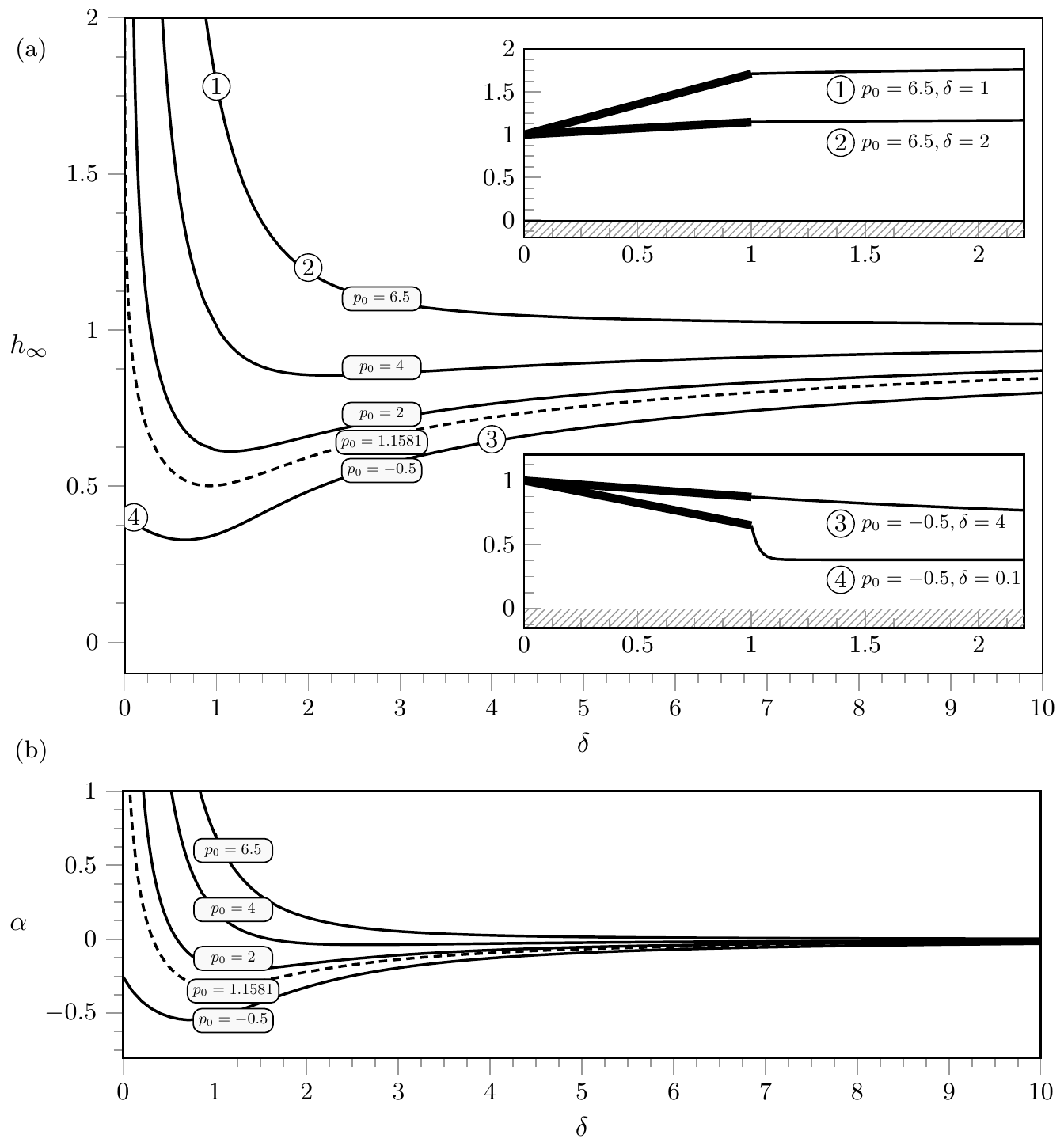}
\caption{
(a) The far-field film height of the pinned problem, $\hinf$, plotted as a function of $\delta$ for five values of the prescribed pressure, $p_0=-0.5$, 1.1581, 2, 4 and 6.
The two insets show four specific solutions, indicated on the main figure by \ding{172} to \ding{175}.
The critical value of $p_0$ that separates the two different behaviours in the limit $\delta \to 0$ is $p_0=\pzerocrit \simeq 1.1581$ (shown with the dashed line).
(b) The tilt angle, $\alpha$, plotted as a function of $\delta$.
}
\label{fig:pinned_hinf}
\end{figure}

Figure \ref{fig:pinned_hinf}(a) shows the far-field film height, $\hinf$, plotted as a function of $\delta$ for five values of the prescribed pressure, $p_0$.
The two insets show four specific solutions, indicated on the main figure by \ding{172} to \ding{175}.
In the limit $\delta \to \infty$, corresponding to slow substrate motion and/or strong surface tension (discussed in \S\ref{sec:pinned_largedelt}), the far-field height always approaches $\hinf=1$, but in the limit $\delta \to 0$, corresponding to fast substrate motion and/or weak surface tension (discussed in \S\ref{sec:pinned_smalldelt}), the far-field height may either approach a finite value (as it does for $p_0=-0.5$) or may tend to infinity (as it does for $p_0=2$, 4 and 6.5).
We will show in \S\ref{sec:pinned_smalldelt} that
the critical value of $p_0$ that separates the two different behaviours in the limit $\delta \to 0$ is $p_0=\pzerocrit \simeq 1.1581$ (shown with the dashed line in Figure \ref{fig:pinned_hinf}).
Figure \ref{fig:pinned_hinf}(b) shows the tilt angle, $\alpha$, plotted as a function of $\delta$; in general, the trends of $\alpha$ are qualitatively similar to the corresponding trends of $\hinf$.

\subsection{Asymptotic analysis for slow substrate motion and/or strong surface tension ($\delta \gg 1$)}
\label{sec:pinned_largedelt}

At leading order in the limit $\delta \to \infty$,
the plate is horizontal, $\alpha=0$, with pressure $p=p_0-6x$ and velocity $u=(1+3z)(1-z)$ underneath it, and
the free surface is flat and at the same height as the pivot, $h \equiv \hinf = 1$.

At higher order, the free surface varies slowly over a long length scale of $O(\delta) \gg 1$.
Introducing an appropriately rescaled horizontal coordinate
defined by $x=1+\delta X$, we seek solutions in the form
\begin{equation}
\alpha =     \sum_{n=1}^\infty \frac{\alpha_n}{\delta^n}, \qquad
\hinf  = 1 + \sum_{n=1}^\infty \frac{h_{\infty n}}{\delta^n}, \qquad
h(x)   = 1 + \sum_{n=1}^\infty \frac{h_n(X)}{\delta^n}.
\label{eq:deltinf_asym}
\end{equation}
Substituting the expansions (\ref{eq:deltinf_asym}) into the Landau--Levich equation (\ref{LandauLevich}) shows that the first-order solution for the free surface, $h_1=h_1(X)$, satisfies
$h_{1XXX}=3(\hinfone-h_1)$
subject to $h_1(0)=\alpha_1$ and $h_1 \to \hinfone$ as $X \to \infty$, and hence is given by
$h_1 = \hinfone + (\alpha_1-\hinfone)\exp\left(-3^{1/3}X\right)$.
Imposing the pressure condition (\ref{p2}) and the moment condition (\ref{M2}) yields
$\alpha_1=0$, $\alpha_2=h_{1X}(0)$ and $h_{1XX}(0)=6-p_0$,
and hence we obtain the solutions
\begin{eqnarray}
\alpha &=&     \frac{ p_0-6 }{3^{1/3}\delta^2}
+ O\left(\frac{1}{\delta^3}\right),
\label{largedeltaalpha} \\
\hinf  &=& 1 + \frac{ p_0-6 }{3^{2/3}\delta}
+ O\left(\frac{1}{\delta^2}\right), \\
\label{largedeltahinf}
h(x)   &=& 1 + \frac{ p_0-6 }{3^{2/3}\delta}
\left[1-\exp\left(-\frac{3^{1/3}x}{\delta}\right)\right]
+ O\left(\frac{1}{\delta^2}\right).
\label{largedeltah}
\end{eqnarray}
In particular, these solutions show that
when $p_0<6$
(\ie when the leading-order pressure at the downstream end of the plate is negative)
the first-order effect of slow substrate motion and/or strong surface tension is to
slightly tilt the plate downwards from the horizontal and to
slightly decrease the height of the free surface.
However,
when $p_0>6$
(\ie when the leading-order pressure at the downstream end of the plate is positive)
the opposite behavior occurs.
In the special case $p_0=6$
it is straightforward to show that
\begin{equation}
\alpha =     \delta^{-3}+O(\delta^{-4}), \qquad
\hinf  = 1 + \delta^{-3}+O(\delta^{-4}), \qquad
h(x)   = 1 + \delta^{-3}+O(\delta^{-4}).
\end{equation}

\subsection{Asymptotic analysis for fast substrate motion and/or weak surface tension ($\delta \ll 1$)}
\label{sec:pinned_smalldelt}

At leading order in the limit $\delta \to 0$,
the plate is tilted at an angle $\alpha=\alpha_0$ and
the free surface is flat and at the height $h\equiv\hinf=\hinfzero$
except in a narrow inner region
near the downstream end of the plate of width $O(\delta) \ll 1$
in which it varies rapidly to match the height of the end of the plate.

\begin{figure}
\includegraphics[width=1.0\textwidth]{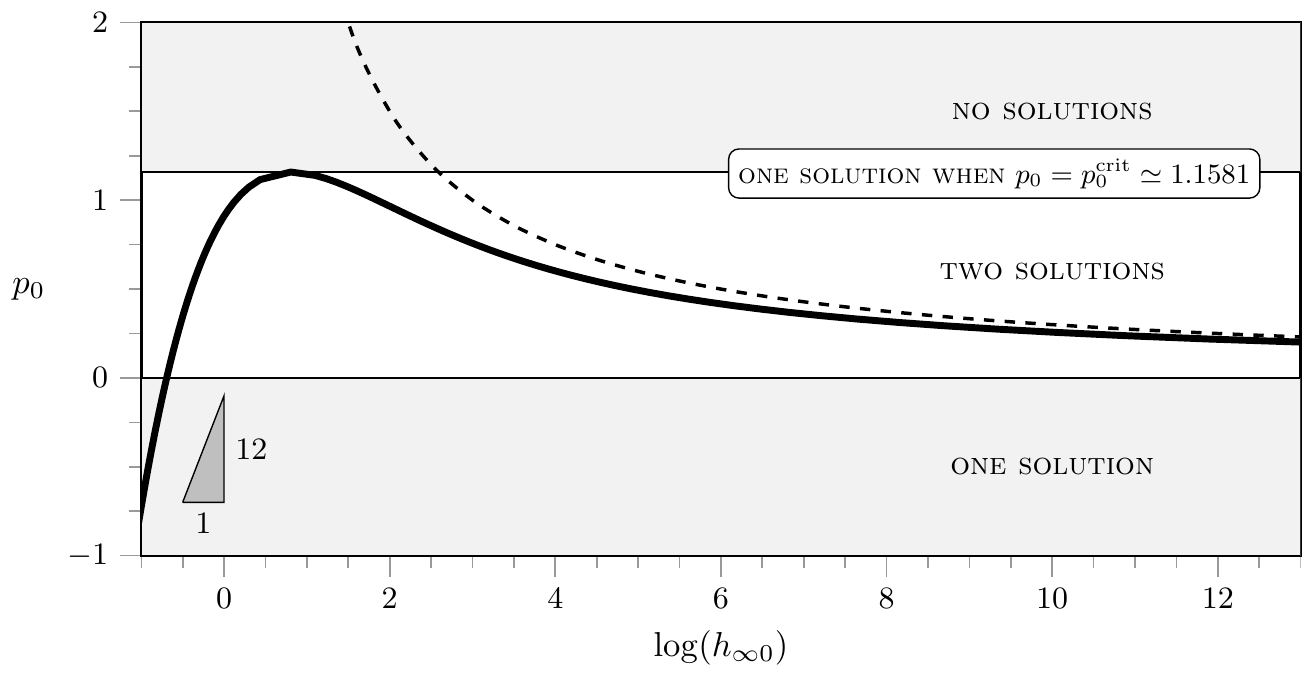}
\caption{
The prescribed pressure, $p_0$, plotted as a function of the logarithm of the leading-order film height, $\log(\hinfzero)$, in the limit $\delta \to 0$, showing that there is
one solution  when $p_0 \le 0$,
two solutions when $0 < p_0 < \pzerocrit \simeq 1.1581$,
one solution  when $p_0 = \pzerocrit$, and
no  solution  when $p_0 > \pzerocrit$.
The dashed line corresponds to the leading-order behaviour in the limit $p_0 \to 0^+$ in which $\hinfzero \to \infty$,
namely $\log\hinfzero = 3/p_0$, and
the triangle shows the slope of the leading-order behaviour in the limit $p_0 \to -\infty$,
namely $\log\hinfzero = p_0/12$.
}
\label{fig:hinge_deltazero}
\end{figure}

Introducing an appropriately rescaled inner horizontal coordinate
defined by $x=1+\delta X$,
the leading-order solution for the free surface in the inner region, $h_0=h_0(X)$,
satisfies the Landau--Levich equation
\begin{equation}
h_{0XXX}=\frac{3(\hinfzero-h_0)}{h_0^3}
\label{LandauLevich1}
\end{equation}
subject to $h_0(0)=1+\alpha_0$ and the matching condition
$h_0 \to \hinfzero$ as $X \to \infty$.
However, this leading-order inner solution is not required in order to determine either
the leading-order tilt angle, $\alpha_0$, or
the leading-order film height, $\hinfzero$.
Specifically, from the leading-order version of the pressure condition (\ref{p2}), $\hinfzero$ is given explicitly by
\begin{equation}
\hinfzero = \frac{(1+\alpha_0)[6+(1+\alpha_0)p_0]}{6(2+\alpha_0)},
\label{hinfzero}
\end{equation}
and $\alpha_0$ can be determined from solving the leading-order version of the moment condition (\ref{M2}), that is, using
\begin{equation}
\frac{p_0}{2} + \int_0^1 \left[ 6 x I_2(x) - 12 \hinf x I_3(x) \right] \de{x} = 0.
\label{M3}
\end{equation}
%
Figure \ref{fig:hinge_deltazero} shows the prescribed pressure, $p_0$, plotted as a function of the logarithm of the leading-order film height, $\log(\hinfzero)$.
In particular, Figure \ref{fig:hinge_deltazero} shows that there is
one solution  when $p_0 \le 0$,
two solutions when $0 < p_0 < \pzerocrit \simeq 1.1581$,
one solution  with $\alpha_0 \simeq 6.9022$ and $\hinfzero \simeq 2.2416$ when $p_0 = \pzerocrit$, and
no  solution  when $p_0 > \pzerocrit$.

Consider first the special case of no prescribed pressure, $p_0=0$.
In this case the plate is horizontal, $\alpha_0=0$, with zero pressure, $p=0$, and simple shear flow, $u=1-z$, underneath it, and the film height is exactly half the height of the plate, $\hinfzero=1/2$.
The Landau--Levich equation (\ref{LandauLevich1}) is solved numerically by shooting from near $h_0=\hinfzero=1/2$ with the appropriate far-field behaviour (\ref{eq:wkbmodes}), and stopping once the value $h_0=1$ is reached.
This provides the leading-order values of the slope and the curvature of the free surface at the end of the plate, with
\begin{subequations}
\begin{gather}
h_{0X}(0)  \sim -0.736 \left(\frac{3^{1/3}}{\hinfzero}\right)   \simeq -2.123,
\label{h0xnum} \\
h_{0XX}(0) \sim  0.424 \left(\frac{3^{1/3}}{\hinfzero}\right)^2 \simeq  3.528.
\label{h0xxnum}
\end{gather}
\end{subequations}
Proceeding to first order we find that
\begin{equation}
\alpha = - \frac{4 \delta h_{0XX}(0)}{3} + O(\delta^2)
\quad \text{and} \quad
\hinf  = \frac{1}{2} - \frac{\delta h_{0XX}(0)}{4} + O(\delta^2).
\label{eq:deltzero_hinflarge}
\end{equation}
In particular, these solutions show that since $h_{0XX}(0) > 0$, the first-order effect of fast substrate motion and/or weak surface tension is to slightly tilt the plate downwards from the horizontal and to slightly decrease the film height from $1/2$.

In the limit of small prescribed pressure, $p_0 \to 0$,
the solution approaches the bounded solution in the case $p_0=0$ according to
\begin{equation}
\alpha_0   \sim \frac{2p_0}{3} \to 0
\quad \text{and} \quad
\hinfzero  \sim \frac{1}{2} + \frac{p_0}{6} \to \frac{1}{2}.
\end{equation}
In the limit of small positive prescribed pressure, $p_0 \to 0^+$,
the other solution becomes unbounded according to
\begin{equation}
\alpha_0  \sim \exp\left(\frac{3}{p_0}+\frac{3}{2}\right) \to \infty
\quad \text{and} \quad
\hinfzero \sim \frac{p_0}{6}\exp\left(\frac{3}{p_0}\right) \to \infty.
\label{eq:delt0_sec}
\end{equation}
The dashed line shown in Figure \ref{fig:hinge_deltazero} corresponds to
the leading-order behaviour of this solution, namely $\log\hinfzero = 3/p_0$.

Finally, in the limit of large negative prescribed pressure, $p_0 \to -\infty$, the plate is sucked towards the substrate by the large negative
pressure generated underneath it and almost all of the fluid is trapped
behind the plate. However, the motion of the substrate generates
a sufficiently large positive pressure near the downstream end
of the plate to prevent the gap closing entirely.
Specifically, $\alpha_0$ and $\hinfzero$ approach $-1$ and zero, respectively, exponentially from above according to
\begin{equation}
\alpha_0  \sim -1 + \exp\left(\frac{p_0}{12}\right) \to -1^+
\quad \text{and} \quad
\hinfzero \sim      \exp\left(\frac{p_0}{12}\right) \to 0^+.
\end{equation}
The triangle shown in Figure \ref{fig:hinge_deltazero} corresponds to
the slope of the leading-order behaviour of this solution, namely $\log\hinfzero = p_0/12$.

Note that, in addition to these bounded (regular) solutions
there is also a unbounded (singular) solution for all $p_0>0$ for which $\hinf \to \infty$ as $\delta \to 0$,
which disappears in the limiting case $\delta=0$.

\subsection{Numerical solutions for general values of $\delta$}
\label{sec:pinned_multiplicity}

\begin{figure}
\includegraphics[width=1.0\textwidth]{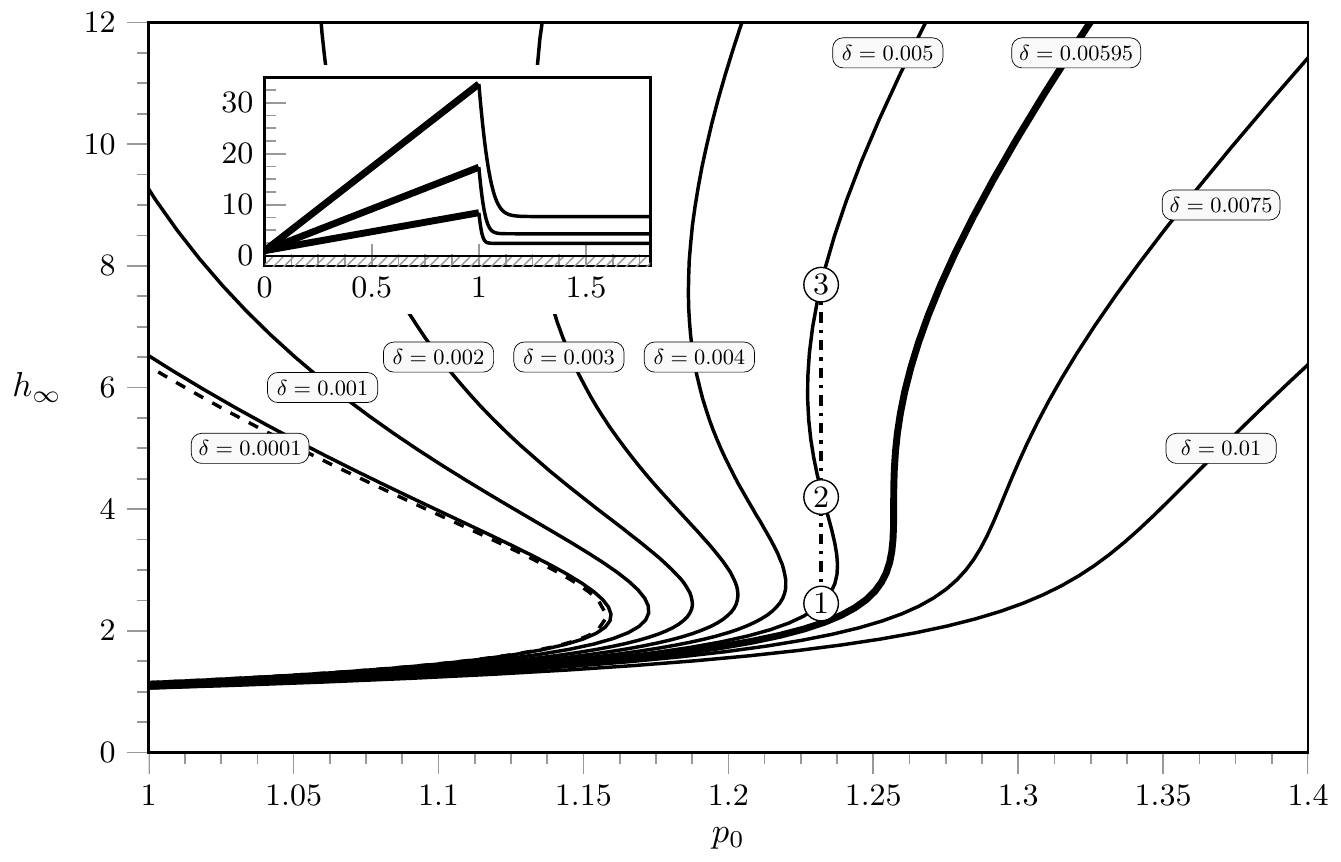}
\caption{
The far-field film height of the pinned problem, $\hinf$, plotted as a function of the prescribed pressure, $p_0$, for various values of $\delta$.
The dashed line corresponds to the bounded asymptotic solution in the limit $\delta \to 0$ described in \S\ref{sec:pinned_smalldelt}, while
the thick line corresponds to the critical solution with $\delta = \deltastar \simeq 0.00595$ at which the three solutions merge into one at a point of vertical tangency at $p_0 = \pzerostar \simeq 1.26$ and $\hinf = \hinfstar \simeq 4.04$.
Note that the $p_0$-axis should extend to negative values (see Figure \ref{fig:hinge_deltazero}), but it has been truncated to better show the details near $\delta = \deltastar$.
The inset shows the three solutions in the case $\delta=0.005 \, (<\deltastar)$ and $p_0=1.232 \, (<\pzerostar)$, indicated on the main figure by \ding{172} to \ding{174}.
}
\label{fig:bif}
\end{figure}

As we have just seen, in the limiting case $\delta=0$ there are zero, one or two bounded solutions of the pinned problem.
In fact, as we shall now describe, numerical solutions for general values of $\delta$ reveal that for all non-zero values of $\delta$ there are between one and three solutions.

Figure \ref{fig:bif} shows the far-field film height, $\hinf$, plotted as a function of the prescribed pressure, $p_0$, for various values of $\delta$ calculated using the numerical method described in \S\ref{sec:num1}.
In the limit $\delta \to 0$ the bounded solutions approach the asymptotic solutions for $p_0 \le \pzerocrit$ described in \S\ref{sec:pinned_smalldelt} (shown with the dashed line) and illustrated in Figure \ref{fig:hinge_deltazero}.
Provided that the value of $\delta$ is not too large (specifically, provided that $\delta < \deltastar$ where the value $\deltastar \simeq 0.00595$ was calculated numerically), then $\hinf$ is an ``S shaped'' function of $p_0$, and so, depending on the value of $p_0$, there are between one and three solutions.
At $\delta = \deltastar$ (shown with the thick line), the three solutions merge into one at a point of vertical tangency at $p_0 = \pzerostar \simeq 1.26$ and $\hinf = \hinfstar \simeq 4.04$, and thereafter for $\delta \ge \deltastar$ then $\hinf$ is a monotonically increasing function of $p_0$ and so there is only one solution for all values of $p_0$.
Similarly,
for $p_0 \ge \pzerostar$ there is only one solution for all values of $\delta$.

The inset in Figure \ref{fig:bif} shows the three solutions in the case $\delta = 0.005 \, (<\deltastar)$ and  $p_0 = 1.232 \, (<\pzerostar)$; they are distinguished by their increasing values of $\alpha$ and $\hinf$.
Although we do not study the stability of such solutions, we conjecture that only the solution that corresponds to the smallest values of $\alpha$ and $\hinf$ are stable (see \citealt{quintans_2009} for the stability analysis for the case of a fixed horizontal plate).

\section{The free-floating problem}
\label{sec:free}

We now turn to the case of flow underneath a free-floating rigid flat plate, as shown in Figure \ref{fig:sketch}(b), that is, the ``contact lens problem'' of \cite{moriarty_1996}.

In this problem we choose $\cH=\hinf$ for the characteristic dimensional film height in (\ref{nondimen});
this sets the non-dimensional far-field film height to be $\hinf=1$.
From (\ref{plateeq}) the non-dimensional height of the plate is given by $H=H_0+\alpha x$, where $\alpha > -H_0$.
As for the pinned problem,
the solutions for the velocity $u$ and pressure $p$ underneath the free surfaces are given by (\ref{u_freesurface_Q}) and (\ref{p_freesurface}), while the corresponding solutions underneath the plate are given by (\ref{u_plate_Q}) and (\ref{p_plate}).
The height of the free surfaces upstream (\ie for $x<0$) and downstream (\ie for $x>1$) of the plate are obtained by solving the Landau--Levich equation (\ref{LandauLevich}) subject to appropriate matching conditions at both ends of the plate and the far-field conditions $h \to 1$ as $\vert x \vert \to \infty$.

\begin{figure}
\includegraphics[width=1.0\textwidth]{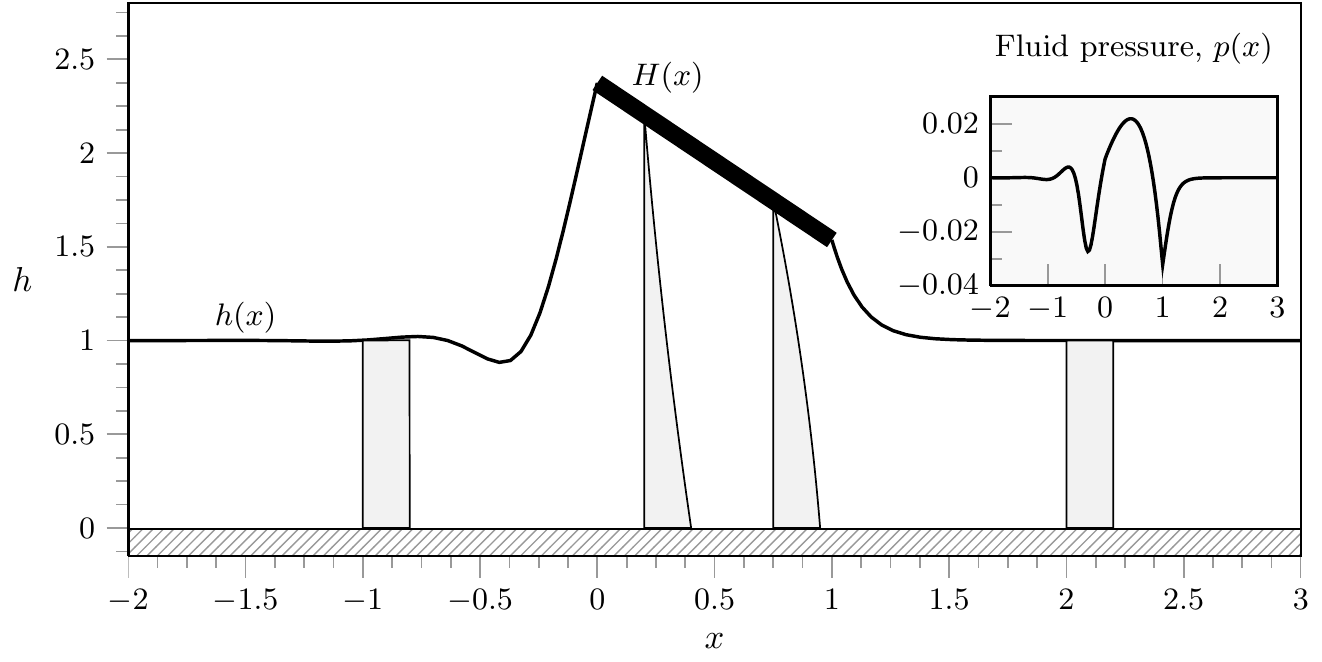}
\caption{
The solution of the free-floating problem when $\delta = 0.104$ for which we find numerically that $\alpha \simeq 0.838$ and $H_0 \simeq 2.374$. Re-scaled velocity profiles, $0.25 \, u(x,z)$, are plotted as functions of $z$ for $x = -1$, 0.2, 0.75, 2. The inset shows the fluid pressure, $p(x)$.
}
\label{fig:free}
\end{figure}

A summary of all the relevant equations and boundary conditions is given in Table \ref{tab:free}, and the solution of the free-floating problem when $\delta=0.104$, computed numerically as described subsequently in \S\ref{sec:num2}, is shown in Figure \ref{fig:free}.
From the earlier discussion concerning (\ref{eq:wkbmodes}), the downstream condition again rules out the two exponentially growing modes. Upstream, however, only the single exponentially growing mode is excluded on physical grounds, so that $C_1 = 0$, but leaving both $C_2$ and $C_3$ to be determined. Thus, in total, the far-field conditions provide three boundary conditions.

\begin{table}
\begin{center}
\begin{tabular}{cp{4cm}p{6cm}}
\qquad Unknowns \qquad \qquad & Equations & Boundary conditions \\
$\alpha$, $H_0$ & $H=H_0+\alpha x$
\newline $\delta^3 h_{xxx}=3(1-h)/h^3$
\newline $p_x=6(H-2)/H^3$ & [1] $h(0)=H_0$
\newline [2] $h(1)=H(1)=H_0+\alpha$
\newline [3] $[p] = 0$ at $x = 0$
\newline [4] $[p] = 0$ at $x = 1$
\newline [5] $h \to \hinf$ as $x \to -\infty$
\newline [6, 7] $h \to \hinf$ as $x \to \infty$
\newline [8] $F_z=0$
\newline [9] $M=0$ \\
\end{tabular}
\end{center}
\caption{
A summary of the equations and boundary conditions of the ninth-order rigid free-floating problem.
}
\label{tab:free}
\end{table}

At both ends of the plate we again assume that the free surface is pinned to the end of plate and
impose continuity of height, namely $h(0)=H(0)=H_0$ and $h(1)=H(1)=H_0+\alpha$, and pressure.
The condition of continuity of pressure at $x=0$ yields
\begin{equation}
p(0) = - \delta^3 h_{xx}(0).
\label{p3}
\end{equation}
Using (\ref{p3})
the condition of continuity of pressure at $x=1$ yields
\begin{equation}
6I_2(1) - 12I_3(1)
+ \delta^3 \left[ h_{xx}(1) - h_{xx}(0) \right] = 0,
\label{p4}
\end{equation}
and hence
\begin{equation}
\frac{6[H_0(H_0+\alpha)-(2H_0+\alpha)]}{H_0^2(H_0+\alpha)^2}
+ \delta^3 \left[ h_{xx}(1) - h_{xx}(0) \right] = 0.
\label{p5}
\end{equation}
The condition (\ref{p5}) explicitly couples the free surfaces upstream and downstream of the plate via the flow underneath it.

The two remaining unknowns, namely the tilt angle, $\alpha$, and the height of the plate at $x = 0$, $H_0$, are determined by imposing zero total vertical force on the plate, $F_z = 0$, and zero total moment about any point of the plate, $M = 0$.

The condition of zero total vertical force on the plate requires that
\begin{equation}
F_z = \int_0^1 p \de{x}
+ \delta^3 \left[h_x(1)-h_x(0)\right] = 0,
\label{F1}
\end{equation}
where the first term represents the vertical force on the plate due to the fluid underneath it, and the second term is the vertical component of the surface-tension forces acting at both ends of the plate.
Substituting the expression for the pressure underneath the plate (\ref{p_plate})
into (\ref{F1}) and using (\ref{p3}) yields
\begin{equation}
F_z = - \delta^3 h_{xx}(0)
+ 6 \int_0^1 \left[I_2(x) - 2 I_3(x)\right] \de{x}
+ \delta^3 \left[h_x(1)-h_x(0)\right] = 0,
\label{F2}
\end{equation}
and hence
\begin{equation}
F_z = - \delta^3 h_{xx}(0)
+ \frac{6}{\alpha^2H_0}\left[\alpha-H_0\log\left(\frac{H_0+\alpha}{H_0}\right)\right]
- \frac{6}{H_0^2(H_0+\alpha)}
+ \delta^3 \left[h_x(1)-h_x(0)\right] = 0.
\label{F3}
\end{equation}

The condition of zero total moment of the forces on the plate about any point on the plate, $x=\xp$ ($0 \le \xp \le 1$), is given by
\begin{equation}
M = \int_0^1 (x-\xp) p \de{x}
+ \delta^3 \left[(1-\xp)h_x(1)+\xp h_x(0)+H(0)-H(1)\right] = 0,
\label{M4}
\end{equation}
where the first term represents the moment of the normal force on the plate due to the fluid underneath it, and the second term is the moment of the normal surface-tension forces acting at both ends of the plate, $x=0$ and $x=1$.
Using the zero total vertical force condition (\ref{F1}) to eliminate
the integral of $p$ appearing in the first term of (\ref{M4}) yields
\begin{equation}
M = \int_0^1 x p \de{x}
+ \delta^3 [h_x(1)+H(0)-H(1)] = 0,
\label{M5}
\end{equation}
and hence
\begin{equation}
M = - \frac{\delta^3 h_{xx}(0)}{2} + \int_0^1 \left[ 6x I_2(x) - 12x I_3(x) \right] \de{x}
+ \delta^3 [h_x(1)-\alpha] = 0,
\label{M6}
\end{equation}
where the integrals of $x I_2$ and $x I_3$ (both of which are functions of $\alpha$ and $H_0$) are given by (\ref{xI2}) and (\ref{xI3}) in the Appendix.

Note that in the special case of a horizontal plate, \ie the special case $\alpha=0$,
equations (\ref{F3}) and (\ref{M6}) reduce to
\begin{equation}
F_z = - \delta^3 h_{xx}(0) + \frac{3(H_0-2)}{H_0^3} + \delta^3 \left[ h_x(1)-h_x(0) \right] = 0
\label{F4}
\end{equation}
and
\begin{equation}
M = - \frac{\delta^3 h_{xx}(0)}{2} + \frac{2(H_0-2)}{H_0^3} + \delta^3 h_x(1) = 0,
\label{M7}
\end{equation}
respectively.
Equation (\ref{F4}) is equivalent to the expression for the total vertical force on the plate given by \cite{moriarty_1996} [their equation (2.9)].
Similarly, equation (\ref{M7}) is equivalent to their expression for the total moment of the force on the plate if the condition $F_z  = 0$ is used [their equation (2.10)].

\subsection{The realistic nature of the present steady-state solutions}
\label{sec:realistic}

At this point, the observant reader may ask: \emph{if the substrate is moving to the right, then what is keeping the plate held in place in the present steady-state solutions?}
For the free-floating plate the total horizontal force on the plate is given by an expression similar to \eqref{Fx_pinned}, namely
\begin{equation}
F_x = -\epsilon \left[ \int_0^1 H_x p  \, \de{x} + \int_0^1 u_z \Bigr\rvert_{z = H} \, \de{x} + \frac{\delta^3}{2} \left\{h_x^2(1) - h_x^2(0)\right\} \right],
\label{F6}
\end{equation}
and is thus of higher order in the aspect ratio $\epsilon$ than the vertical component $F_z$ given by (\ref{F3}). Nevertheless, $F_x$ will be non-zero, and hence the steady-state contact lens problem of \cite{moriarty_1996} does not provide a proper balance of horizontal forces, and so in reality, the plate would be swept downstream by the flow. Thus, as we have posed it, the free-floating problem must be accompanied by a small imposed horizontal force, which holds the plate in place horizontally but allows its vertical position to be determined by the flow.

There is, however, a less artificial interpretation of our results for the free-floating problem in a context in which the parameter $\delta$ is associated with time dependence.
From (\ref{F6}) the dimensional drag per unit width per unit length of the plate is $O({\mu U}/{\epsilon L})$, and hence if we consider a plate of dimensional mass per unit width, $m$, moving over a stationary substrate at a non-constant dimensional speed $U=U(t)$ then, by Newton's second law,
\begin{equation}
m \frac{\de U}{\de t} \sim - \frac{\mu U}{\epsilon}.
\label{newton}
\end{equation}
Thus, we can interpret the present ``steady-state'' solutions as snapshots in time where the speed $U=U(t)$, and hence the non-dimensional parameter
$\delta=\delta(t)=(\gamma \epsilon^3)(\mu U(t))$,
is not prescribed, but is rather evolved using (\ref{newton}).
A sequence of solutions for increasing values of $\delta$ then corresponds to a quasi-steady time evolution of a plate as it ultimately comes to rest (\ie to speed $U=0$) in the limit $\delta\to\infty$.

\subsection{Numerical method for the free-floating problem}
\label{sec:num2}

Similarly to the pinned problem,
for general values of $\delta$, the free-floating problem can be solved numerically by reposing it as a nested shooting problem; now, however, we must deal with the added difficulty of matching the upstream and downstream flows.

For each value of $\delta$, begin by guessing an initial value of the tilt angle,
\begin{equation}
\alpha \simeq \alpha_\text{guess},
\end{equation}
and the height of the upstream end of the plate,
\begin{equation}
H_0 \simeq H_{0,\text{guess}}.
\end{equation}

By writing $h = 1 + \kappa$, where $\kappa$ is a numerically small perturbation ($\kappa = 10^{-8}$ in our calculations), the far-field behaviour (\ref{eq:wkbmodes}) is used to provide the values of $h'$ and $h''$. With these three initial conditions, the Landau--Levich equation (\ref{LandauLevich}) is solved for the film height, $h$, until reaching the value of $h = 1 + \alpha_\text{guess}$ at the downstream end of the plate.

Next, the correct value of $\alpha$ (albeit for an incorrect value of $H_0$) is found by applying the zero moment condition (\ref{M6}), and thereafter, both the height and the curvature of the free surface at the upstream end of the plate are ``known''.
As was demonstrated in \cite{moriarty_1996} and \cite{mcleod_1996} (see also \citealt{tuck_1990}), for each pair of values of $h(0)$ and $h_{xx}(0)$ there exists zero, one or two possible values of $h_x(0)$ such that the free surface tends to the constant value $h=1$ as $x \to -\infty$.
To be more specific, and to remove the $\delta$ dependence, let us write
$x = - C s$ where $C = 3^{-1/3} \delta$.
This transforms the Landau--Levich equation (\ref{LandauLevich}) into
\begin{equation}
h_{sss} = \frac{h-1}{h^3},
\label{govfluid2switch}
\end{equation}
where $s > 0$ corresponds to upstream and $s < -1$ to downstream.
For a given pair of values of $h(0)$ and $h_{ss}(0)$, let us write $h_s^*$ for a correct value of $h_s(0)$ (if it exists), such that $h \to 1$ as $s \to \infty$.
If the Landau--Levich equation (\ref{govfluid2switch}) is solved with $h_s(0) > h_s^*$, then the solution blows up; if instead it is solved with $h_s(0) < h_s^*$, then the solution intersects the substrate at a finite value of $s$.

\begin{figure}
\includegraphics[width=1.0\textwidth]{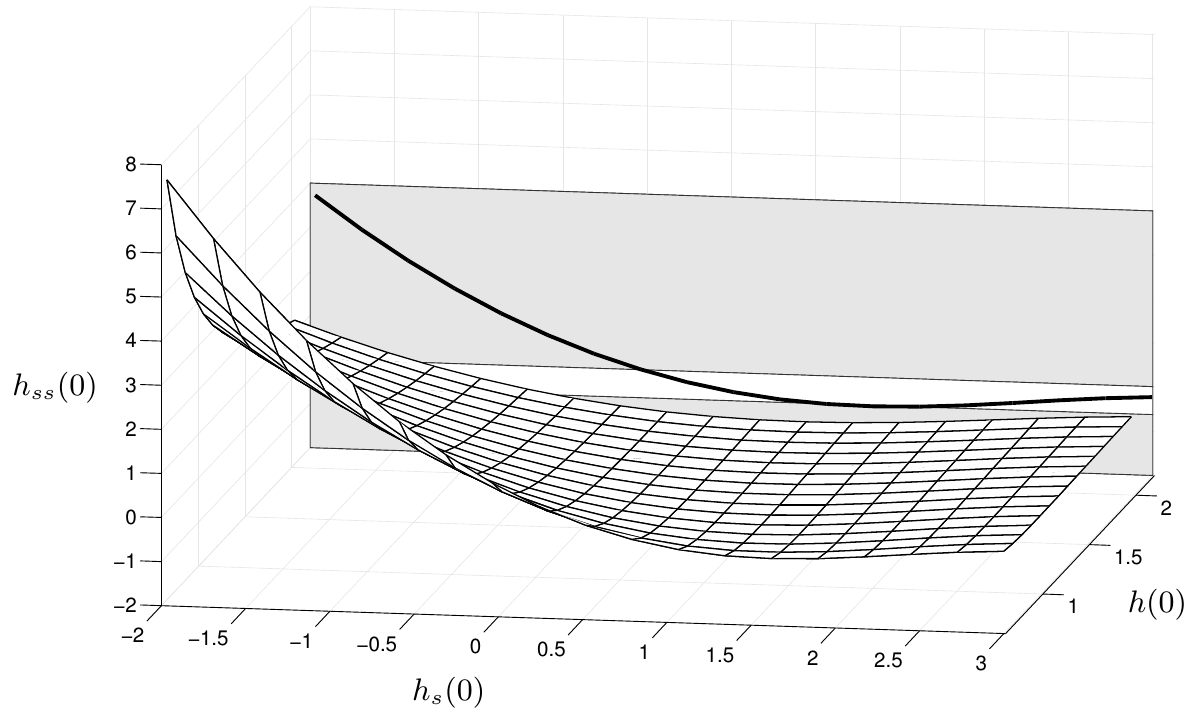}
\caption{
Plot of $h_{ss}(0)$ as a function of $h(0)$ and $h_s(0)$ for which solutions of the Landau--Levich equation (\ref{govfluid2switch}) tend to the constant value $h=1$ as $s \to \infty$. Projected onto the plane is the solution corresponding to a fixed value, $h(0) = 1$, and on this plane the gray-white-gray zones delineate, from top to bottom, regions where there is one solution, two solutions, and no solutions.
}
\label{Hppmat}
\end{figure}

The process of selecting the correct value (or values) of $h_s(0) = h_s^*$ can be automated by using a bisection method as follows.
We specify an interval $[0,s_\text{max}]$, where $s_\text{max}$ is sufficiently large and solve (\ref{govfluid2switch}), searching for an initial lower bound $h_s^\text{lower}$ (above which the solution intersects the substrate), and for an initial upper bound $h_s^\text{upper}$ (below which the solution tends to infinity). Equation (\ref{govfluid2switch}) is then solved with $h_s(0) = (h_s^\text{lower}+h_s^\text{upper})/2$, and depending on its behaviour, we assign this new value of $h_s(0)$ to either $h_s^\text{lower}$ or $h_s^\text{upper}$; the process is repeated until the bisection interval is smaller than a certain tolerance ($10^{-8}$ in our calculations).
Figure \ref{Hppmat} shows the three-dimensional surface formed by the triplet of values $\{h(0), h_s(0), h_{ss}(0)\}$.
Note that although we can generally use interpolated values of the surface to provide an initial guess of the correct shooting value, the sensitivity of the boundary value problem requires that, in order to achieve a reasonable degree of accuracy, the bisection method be applied for each new numerical calculation.

Once the correct value of $h_x(0)$ has been computed, then the zero vertical force condition (\ref{F3}) provides us with the error in using the incorrect value of $H_{0,\text{guess}}$. A nonlinear solver (such as Newton's method) is then applied to find the correct values of $\alpha$ and $H_0$. The major difficulty of the free-floating problem over that of the pinned problem described in \S\ref{sec:num1}, is the necessity of automating the numerical solution for the upstream problem.
The solution of the free-floating problem when $\delta = 0.104$ was shown in Figure \ref{fig:free}.

\begin{figure}
\includegraphics[width=1.0\textwidth]{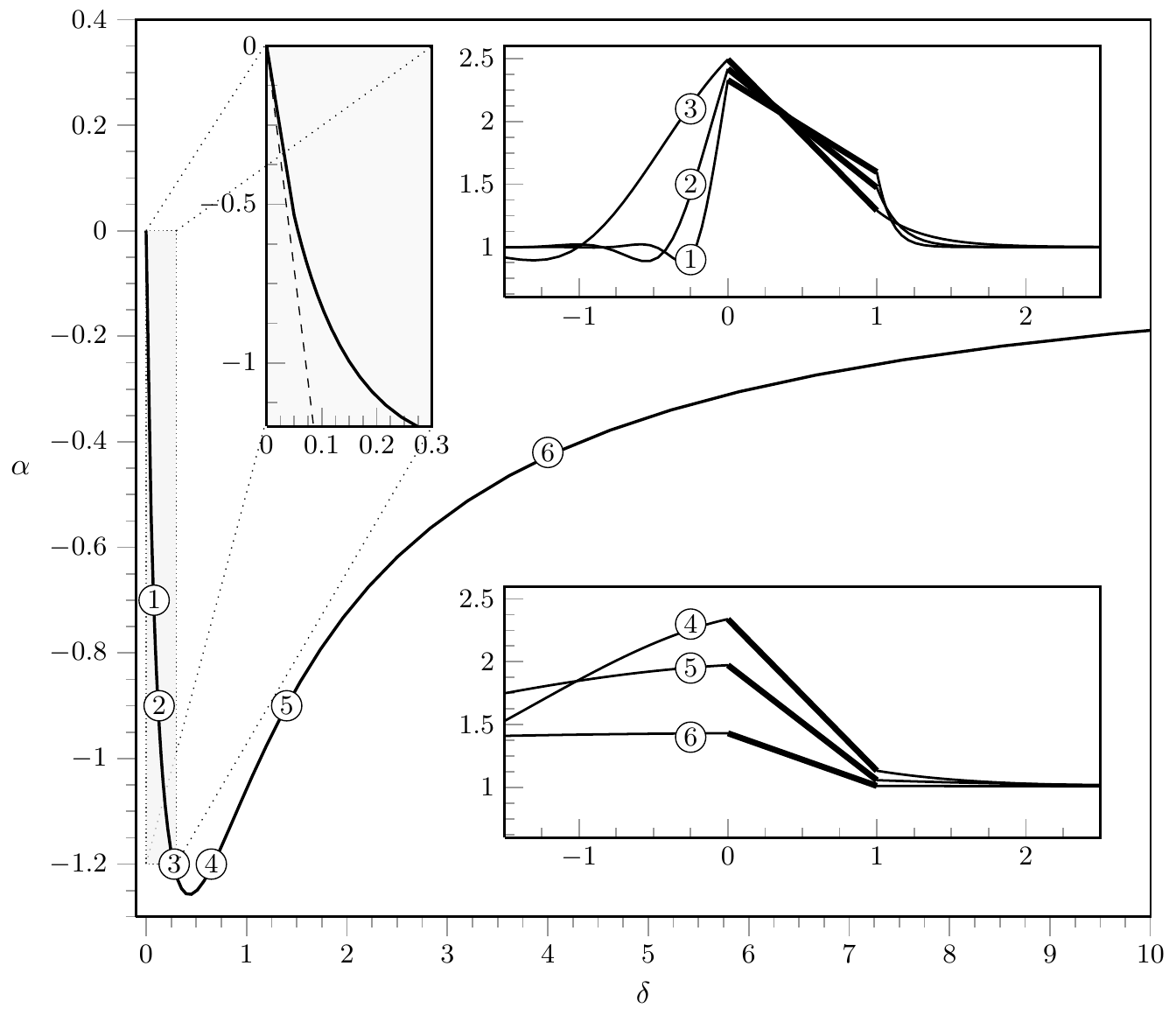}
\caption{
The tilt angle of the free-floating problem, $\alpha$, plotted as a function of $\delta$. The two rightmost insets show six specific solutions, indicated on the main figure by \ding{172} to \ding{177}. The leftmost inset is an enlargement of the grey region in which the dashed line corresponds to the leading-order behaviour in the limit $\delta \to 0$, namely $\alpha = -14.1 \, \delta$.
}
\label{fig:alpha}
\end{figure}

Figure \ref{fig:alpha} shows the tilt angle, $\alpha$, plotted as a function of $\delta$, and the two rightmost insets show six specific solutions, indicated on the main figure by \ding{172} to \ding{177}.
The tilt angle tends to zero in the limits $\delta \to 0$ and $\delta \to \infty$, showing that a horizontal plate satisfies the equilibrium conditions of zero vertical force and zero moment in these two extreme situations (as was originally claimed by \citealt{moriarty_1996}), but for any finite and non-zero value of $\delta$, the plate must tilt and move upwards or downwards in order to satisfy the equilibrium conditions.
The plot for the height of the upstream end of the plate, $H_0$, as a function of $\delta$ (not shown for brevity) shows that the trends of $H_0$ are qualitatively similar to the corresponding trends of $\alpha$, but with $H_0=2$ at $\delta=0$, increasing to a maximum initially as $\delta$ increases, and then decreasing again to $H_0=2$ from above as $\delta \to \infty$.

The intrinsic difficulty of the shooting procedure for the free-floating problem lies in the fact that,
for a given position of the upstream end of the plate, $h(0)$,
the combinations of possible upstream free-surface slopes, $h_s(0)$, and curvatures, $h_{ss}(0)$,
depend on where one lies on the surface shown in Figure \ref{Hppmat}.
Specifically,
for a given value of $h_{ss}(0)$ there may be either zero, one or two corresponding values of $h_s(0)$.
Non-unique solutions might be possible; for example,
if there exists a free-surface configuration with the same upstream height, $h(0)$, and curvature, $h_{ss}(0)$,
but two valid values of the upstream slope, $h_s(0)$,
for which the free surface tends to the constant value $h=1$ upstream.
However,
the numerical scheme is extremely complicated because the triplet of values
$\{h(0)$, $h_s(0)$, $h_{ss}(0)\}$
must then be linked to the downstream flow, and then to the conditions of zero total force and moment.
It can be verified that all of the present numerical solutions lie
to the left of the minimum in the $\{h_s(0), h_{ss}(0)\}$ graphs.
However, we cannot be certain whether or not other solutions exist,
and searching for them numerically does not seem to be possible using the current numerical scheme.

\subsection{Asymptotic analysis for slow substrate motion and/or strong surface tension ($\delta \gg 1$)}

At leading order in the limit $\delta \to \infty$,
the plate is horizontal, $\alpha=0$, and at the same height as the film, $H_0=1$, and
the free surface is flat, $h \equiv \hinf = 1$.

At higher order, the free surfaces upstream and downstream of the plate both vary slowly over a long length scale of $O(\delta) \gg 1$. Thus we seek solutions for $\alpha$ and $H_0$ in the form
\begin{equation}
\alpha =     \sum_{n=1}^\infty \frac{\alpha_n}{\delta^n}, \quad
H_0    = 1 + \sum_{n=1}^\infty \frac{H_{0n}}{\delta^n}.
\end{equation}
Writing $x=1+\delta X$ and
seeking a solution for the downstream free surface in the form
\begin{equation}
h(x) = 1 + \sum_{n=1}^\infty \frac{h_n(X)}{\delta^n}
\end{equation}
shows that $h_1=h_1(X)$ satisfies
$h_{1XXX}=-3h_1$
subject to $h_1(0) = H_{01} + \alpha_1$ and $h_1 \to 0$ as $X \to \infty$,
and hence is given by
\begin{equation}
h_1=(H_{01}+\alpha_1)\exp\left(-3^{1/3}X\right).
\end{equation}
Similarly, writing $x=\delta \Xhat$ and seeking a solution for the upstream free surface in the form
\begin{equation}
h(x) = 1 + \sum_{n=1}^\infty \frac{\hhat_n(\Xhat)}{\delta^n}
\end{equation}
shows that $\hhat_1=\hhat_1(\Xhat)$ satisfies
$\hhat_{1\Xhat\Xhat\Xhat}=-3\hhat_1$
subject to $\hhat_1(0)=H_{01}$ and $\hhat_1 \to 0$ as $\Xhat \to -\infty$,
and hence is given by
\begin{equation}
\hhat_1(\Xhat) = \frac{H_{01}}{\cos\phi}\exp\left(\frac{3^{1/3}\Xhat}{2}\right)
\cos\left(\frac{3^{5/6}\Xhat}{2}+\phi\right),
\end{equation}
where the unknown phase angle $\phi$ is to be determined.
Imposing (\ref{p5}), (\ref{F3}) and (\ref{M5}) yields
\begin{equation}
\alpha_1=0, \qquad
H_{01}=\frac{2}{3^{2/3}} \simeq 0.9615, \qquad \phi=\frac{\pi}{3},
\end{equation}
and hence we obtain the solutions
\begin{gather}
\alpha=O\left(\frac{1}{\delta^2}\right), \\
H_0=\frac{2}{3^{2/3}\delta} + O\left(\frac{1}{\delta^2}\right), \\
h(x)=1 + \left\{
\begin{array}{ll}
\displaystyle
\frac{2}{3^{2/3}\delta}
\exp\left(-\frac{3^{1/3}(x-1)}{\delta}\right)
& \hbox{for} \quad x \ge 1 \\
\displaystyle
\frac{4}{3^{2/3}\delta}
\exp\left(\frac{3^{1/3}x}{2\delta}\right)
\cos\left(\frac{3^{5/6}x}{2\delta}+\frac{\pi}{3}\right)
& \hbox{for} \quad x \le 0
\end{array}
\right\} + O\left(\frac{1}{\delta^2}\right).
\end{gather}

\subsection{Asymptotic analysis for fast substrate motion and/or weak surface tension ($\delta \ll 1$)}

At leading order in the limit $\delta \to 0$,
the plate is horizontal, $\alpha=0$, and at twice the height of the film, $H_0=2$, and
the free surface is flat, $h \equiv \hinf = 1$,
except in narrow inner regions near the upstream and downstream ends of the plate of width $O(\delta) \ll 1$.

Writing $x=1+\delta X$ the leading-order inner solution
for the downstream free surface, $h=h_0(X)$, satisfies the Landau--Levich equation
(\ref{LandauLevich1}) with $\hinfzero$ set equal to unity
subject to $h_0(0) = 2$ and $h_0 \to 1$ as $X \to \infty$. In particular, solving this problem numerically yields the values $h_{0X}(0) \simeq -1.06$ and $h_{0XX}(0) \simeq 0.882$.

Similarly, writing $x=\delta \Xhat$ the leading-order inner solution
for the upstream free surface, $\hhat_0=\hhat_0(\Xhat)$, satisfies the Landau--Levich equation
(\ref{LandauLevich1}) with $\hinfzero$ set equal to unity and $X$ and $h_0$ replaced by $\Xhat$ and $\hhat_0$, respectively,
subject to $\hhat_0(0)=2$ and $\hhat_0 \to 1$ as $\Xhat \to -\infty$.
Solving this problem numerically requires one additional boundary condition, and this is obtained by seeking the first-order-accurate solution for position of the plate in the form
$\alpha =     \delta \alpha_1 + O(\delta^2)$
and
$H_0    = 2 + \delta H_{01} + O(\delta^2)$.
Imposing (\ref{p5}), (\ref{F3}) and (\ref{M5}) yields
\begin{equation}
\alpha_1 = -16h_{0XX} , \quad
H_{01} = 8h_{0XX}, \quad
\hhat_{0\Xhat\Xhat}(0)=h_{0XX}(0),
\end{equation}
the third of which provides the additional boundary condition.
Hence we obtain the solutions
\begin{equation}
\alpha=-14.1\,\delta+O(\delta^2)
\quad \text{and} \quad
H_0 = 2 + 7.06\,\delta+O(\delta^2).
\end{equation}
The dashed line shown in the leftmost inset in Figure \ref{fig:alpha} corresponds to the leading-order behaviour in the limit $\delta \to 0$, namely $\alpha = -14.1 \delta$.
Finally, solving numerically for the leading-order inner solution for the upstream free surface yields the value $\hhat_{0\Xhat}(0) \simeq 1.59$.

\section{Discussion}
\label{sec:discussion}

In the present work we investigated two paradigm problems for a range of fluid-structure interaction problems in which viscosity and surface tension both play an important role, namely a pinned or free-floating rigid plate lying on the free surface of a thin film of viscous fluid, which itself lies on top of a horizontal substrate that is moving to the right at a constant speed.

For the pinned problem we showed that,
depending on the values of the inverse capillary number, $\delta$, and the prescribed pressure, $p_0$,
there are between one and three solutions with non-zero tilt angle, with a unique solution
for all $p_0$ for sufficiently large values of $\delta \ge \deltastar$
and
for all $\delta$ for sufficiently large values of $p_0 \ge \pzerostar$.
For the free-floating problem we showed that
there is a solution with non-zero tilt angle for all values of $\delta$.
However, the question of whether or not other solutions exist
remains a challenging open problem.
In both problems we showed that,
in contradiction to the conjecture by \cite{moriarty_1996} for the free-floating problem,
the plate has to tilt in order satisfy the equilibrium conditions of zero vertical force and zero moment, and
only becomes horizontal in the limit $\delta \to \infty$ and in the limit $\delta \to 0$ for the free-floating problem.

One of the main conclusions of the present
work is that even the deceptively simple problem of a rigid plate moving relative to a rigid substrate has considerable
difficulties and subtleties which must be understood before elastic effects can be fully understood.
The crux of the problem is the study of the Landau--Levich equation (\ref{LandauLevich}) which, while simple in appearance, contains many hidden complications.
For example, we observed in \S\ref{sec:free} that in the upstream part of the flow, the multiplicity of solutions makes it difficult to design a robust numerical routine, especially when such a routine must automate the matching between upstream and downstream flows in the free-floating problem.
So while the numerical computations were used to verify the asymptotic solutions, in many cases, obtaining a convergent numerical result demanded that we begin with the asymptotic solutions as an initial guess.
The difficulty of numerically computing the boundary value problems summarized in Tables \ref{tab:pinned} and \ref{tab:free} makes obtaining the asymptotic solutions a pragmatic necessity, rather than just a purely theoretical exercise.

In Part 2 of the present study we will build on the foundations laid in Part 1 to address the case of an elastic plate which provided much of the original motivation for this work.
Mathematically, this requires coupling the third-order Landau--Levich equation for the location of the free surface to a fifth-order (Euler-Bernoulli) equation for the location of the plate, and introduces additional complications to obtaining numerical solutions to the resulting coupled problem.

\begin{acknowledgements}
The authors thank Drs Peter Howell and Dominic Vella (University of Oxford) for valuable discussions.
This work was begun while SKW was a Visiting Fellow in the Department of Mechanical and Aerospace Engineering at Princeton University, and completed while SKW was a Visiting Fellow and PHT was a Short Term Visitor at the Oxford Centre for Collaborative Applied Mathematics (OCCAM) at the University of Oxford.
This publication was based on work supported in part by Award No KUK-C1-013-04, made by King Abdullah University of Science and Technology (KAUST).
SKW is presently a Leverhulme Trust Research Fellow (2013--2015) supported by award RF-2013-355.
HAS acknowledges partial support from NSF grant CBET 1132835.
\end{acknowledgements}

\appendix
\section{Explicit expressions for $I_2$, $I_3$ and their integrals}

With the height of the plate defined by (\ref{plateeq}),
namely $H(x) = H_0 + \alpha x$,
we have
\begin{subequations}
\begin{align}
I_2(x) = \int_0^x \frac{1}{H^2(\tildex)} \de{\tildex} &
= \frac{x}{H_0(H_0 + \alpha x)},
\label{I2} \\
I_3(x) = \int_0^x \frac{1}{H^3(\tildex)} \de{\tildex} &
= \frac{x(2H_0 + \alpha x)}{2H_0^2(H_0 + \alpha x)^2},
\label{I3}
\end{align}
\end{subequations}
and hence
\begin{subequations}
\begin{align}
\int_0^1 I_2(x) \de{x} &
= \frac{1}{\alpha^2 H_0}
\left[\alpha - H_0\log\left(\frac{H_0+\alpha}{H_0}\right)\right],
\label{I2int} \\
\int_0^1 I_3(x) \de{x} &
= \frac{1}{2H_0^2(H_0+\alpha)},
\label{I3int} \\
\int_0^1 xI_2(x) \de{x} &
= \frac{1}{2\alpha^3H_0}
\left[\alpha(\alpha-2H_0)
+ 2H_0^2\log\left(\frac{H_0+\alpha}{H_0}\right) \right],
\label{xI2} \\
\int_0^1 xI_3(x) \de{x} &
= \frac{1}{4\alpha^3H_0^2}
\left[ \frac{\alpha\left\{\alpha(H_0+\alpha)+2H_0^2\right\}}{H_0+\alpha}
- 2H_0^2\log\left(\frac{H_0+\alpha}{H_0}\right) \right].
\label{xI3}
\end{align}
\end{subequations}
Note that for the pinned problem,
we set $H_0=1$ and leave $\hinf$ unknown,
but for the free-floating problem,
we set $\hinf=1$ and leave $H_0$ unknown.



\providecommand{\noopsort}[1]{}

\end{document}